%% file: main.tex
\documentclass[graybox]{svmult}

\usepackage{mathptmx}       
\usepackage{helvet}         
\usepackage{courier}        
\usepackage{type1cm}        
\usepackage{makeidx}         
\usepackage{graphicx}        
\usepackage{multicol}        
\usepackage[bottom]{footmisc}
\usepackage{tcolorbox}
\usepackage{float}
\usepackage{multirow}
\usepackage{subfigure}
\usepackage[T1]{fontenc}
\usepackage[nomargin,inline,marginclue,draft]{fixme}
\usepackage[font=small,labelfont=bf,tableposition=top]{caption}

\makeindex             

\begin{document}

\title*{Birds of a Feather Flock Together? A Study of Developers' Flocking and Migration Behavior in GitHub and Stack Overflow}

\author{Michael Mu Sun, Akash Ghosh, Rajesh Sharma, and Sandeep Kaur Kuttal}

\institute{Michael Mu Sun, Akash Ghosh, Sandeep Kaur Kuttal \at University of Tulsa, Tulsa, \email{sun@utulsa.edu, akashghosh@utulsa.edu, sandeep-kuttal@utulsa.edu}
\and Rajesh Sharma \at University of Tartu, Estonia \email{rajesh.sharma@ut.ee}}
\authorrunning{M. M. Sun, A. Ghosh, R. Sharma and S. K. Kuttal}
\titlerunning{M. M. Sun, A. Ghosh, R. Sharma and S. K. Kuttal}
\maketitle


\input{abstract.tex}
\input{intro.tex}
\input{Background.tex}
\input{relatedwork.tex}

\input{methodology.tex}
\input{results.tex}
\input{disscon.tex}
\input{threat.tex}
\input{conclusion.tex}
\section*{ACKNOWLEDGMENTS }
This work is supported by H2020 framework project, SoBigData, grant no. 654024. 
\bibliographystyle{spmpsci}
\bibliography{reference.bib}

\end{document}

%% file: abstract.tex
\abstract{Interactions between individuals and their participation in community activities are governed by how individuals identify themselves with their peers. We want to investigate such behavior for developers while they are learning and contributing on socially collaborative environments, specifically code hosting sites and question/answer sites. In this study, we investigate the following questions about advocates, developers who can be identified as well-rounded community contributors and active learners. Do advocates flock together in a community? How do flocks of advocates migrate within a community? Do these flocks of advocates migrate beyond a single community? To understand such behavior, we collected 12,578 common advocates across a code hosting site - GitHub and a question/answering site - Stack Overflow. These advocates were involved in 1,549 projects on GitHub and were actively asking 114,569 questions and responding 408,858 answers and 1,001,125 comments on Stack Overflow. We performed an in-depth empirical analysis using social networks to find the flocks of advocates and their migratory pattern on GitHub, Stack Overflow, and across both communities. We found that 7.5\% of the advocates create flocks on GitHub and 8.7\% on Stack Overflow. Further, these flocks of advocates migrate on an average of 5 times on GitHub and 2 times on Stack Overflow. In particular, advocates in flocks of two migrate more frequently than larger flocks. However, this migration behavior was only common within a single community. Our findings indicate that advocates' flocking and migration behavior differs substantially from the ones found in other social environments. This suggests a need to investigate the factors that demotivate the flocking and migration behavior of advocates and ways to enhance and integrate support for such behavior in collaborative software tools.}

%% file: intro.tex
\section{Introduction}
\def\abbre#1{\expandafter\abbreA#1 \relax/ }
\def\abbreA#1#2 {\ifx#1\relax \else\ifnum\uccode`#1=`#1#1\fi\expandafter\abbreA\fi}


``Byrdes of on kynde and color flok and flye allwayes together.'' 
William Turner introduced this phrase, which highlights that individual interactions with other peers in the community is primarily governed by their identification with others and influences their participation in community activities 
\cite{Subramaniam2012,Gershenson2017,Ford2017a}. Researchers from different backgrounds have found compelling evidence of this phenomena in the domain of politics \cite{Anders2013}, scientific research \cite{Tang2013}, medical tests \cite{Vieira2013}, friendship among adolescents \cite{Hamm2000} and email spammers \cite{Yehonatan2015}. However, there exists little knowledge on whether this phenomena of flocking and migration is true for software developers. Without such knowledge, tool builders and researchers might make wrong assumptions about developers' true needs for online peer production sites.
In this paper, we investigated the flocking and migration behavior among developers using Git Hub (GH) and Stack Overflow (SO), two peer production sites popular with the software developer community. \abbre{Git Hub} and \abbre{Stack Overflow} help software developers display their technical and social skills. \abbre{Git Hub} has been the largest open source platform for code hosting and version control. It has more than 1.8M businesses and organizations, 27M developers worldwide, and 80M repos\cite{githubdata}.
\abbre{Stack Overflow} is currently the largest online community for developers to build their careers by learning-sharing programming knowledge via Question/Answer. It contains over 14M questions and 19M answers, hosts an avg. of 51K developers at any given moment and 50M visitors each month \cite{sovdata}.

A competent developer has to be exceptional in technical as well as social skills. Code hosting sites like \abbre{Git Hub} display developers' passion for developing software while technical Question/Answer sites like \abbre{Stack Overflow} reflect their altruistic nature to help the community to learn. The developers who are active in both communities can be identified as altruistic, hobbyist and protean coders. Further, Lee et al.\cite{Lee2017} found that the developers share common interests when they co-commit or do co-pull-requests in the same \abbre{Git Hub} repositories and answer the same \abbre{Stack Overflow} questions. Hence, we study these developers who are active in both \abbre{Git Hub} and \abbre{Stack Overflow} sites and call them as \textbf{``advocates''} throughout our paper.  

To understand the flocking and migration of advocates active on both peer production sites, we formulated following research questions:
\begin{itemize}
	\item RQ1: Do advocates flock on a peer production site?
		\begin{itemize}
        	\item How do the advocates tend to flock? 
            \item What characteristics motivated advocates to flock?
        \end{itemize}
	\item RQ2: How do flocks of advocates migrate within a peer production site?
		\begin{itemize}
        	\item How do the advocates migrate with-in the sites?  
            \item What characteristics motivate a flock of advocates to migrate?
        \end{itemize}
	\item RQ3: Do the flocks of advocates migrate beyond a single peer production site?
		\begin{itemize}
        	\item What characteristics motivate a flock to migrate across platforms?
          \end{itemize}	
\end{itemize}

To answer the above research questions, we collected 12,578 common advocates across a code hosting site - \abbre{Git Hub} - and a question/answering site - \abbre{Stack Overflow}. These advocates were involved in 1,549 projects on \abbre{Git Hub} and were actively asking 114,569 questions and responding with 408,858 answers and 1,001,125 comments on \abbre{Stack Overflow}. On doing an in-depth empirical analysis using social network sciences, we found that 8\% of the advocates create flocks on \abbre{Git Hub} and 9\% on \abbre{Stack Overflow}. Further, these flocks of advocates migrate on an average of 5 times on \abbre{Git Hub} and 2 times on \abbre{Stack Overflow}.  In particular, advocates in flocks of two migrate more frequently than larger flocks. However, this migration behavior was only common within a single community.

The rest of the paper is organized as follows. In Section~\ref{sec:related}, we present related work on social networks and software engineering related to migration and flocking. In Section~\ref{sec:method}, we describe our dataset and in Section~\ref{sec:results} we present results to our research investigations. Section~\ref{sec:concl} concludes the work by also discussing the implications of the results.

%% file: Background.tex
\section{Background of Social Network}
\label{sec:2}

The field of Social networks is based on graph theory, which treats individuals as nodes and the connections among these individuals as edges \cite{SNA}. Let $G(N,E)$ be an undirected graph representing the $nodes (N)$ as users and $edges (E)$ as their interactions in a platform, where $N$ represents all the users of a platform under investigation, and E as a set of interactions among these users. Two users $n_i$ and $n_j$ in a Graph $G$, are connected if they have interacted on the social platform. Next, we explain social network terminologies:

\begin{itemize}
\item \textbf{Degree: }Let $\aleph_i$ represents a set of nodes connected to a node $n_i$ $\in$ $N$. Thus, |$\aleph_i$|, representing the cardinality of the set $\aleph_i$ is the degree of the the node $n_i$. In common terms, each connected node in the set $\aleph_i$ is also called as neighbor of the node $n_i$. The average degree of all the nodes in the network is termed as average degree of the network and is represented as 1/$|N|$$\sum_{i=1}^{|N|}$$\aleph_i$.

\item \textbf{Path: }The path between any two nodes $n_i$ and $n_j$ $\in$ $N$ is represented by $P$($n_i$, $n_j$) = $<${$n_i$, $n_{i+1}$, $n_{i+2}$, ... $n_{j-1}$,$n_j$}$>$, which is a distinct sequence of nodes connected with each other. The cardinality of the path is called as distance. It is possible that there might exist many paths between any two nodes. The path with the \textbf{shortest distance} between any two nodes, is a path set which contains minimum number of nodes among all the paths set.
\item \textbf{Diameter: }It is the cardinality of the greatest path among all the shortest paths for all pair of nodes.

\item \textbf{Average Path Length (APL): }It is the average of all the shortest distances for all pair of nodes. Let $D_{All}$ represents the set of total shortest paths among all pair of nodes. Let $d$($n_i$, $n_j$) represents the shortest distance between two nodes $n_i$, $n_j$ $\in$ $N$, then average path length can be represented as 1/$|D_{All}|$$\sum_{i,j \in N \& i \neq j}^{|N|}$$d(n_i, n_j)$. Average path length provides information about how distant or far any two nodes are present in the network on an average. Diameter provides information about the maximum distance among all pairs of nodes. Thus, these two metrics can be used to infer how stretched or spread network is.

\item \textbf{Edge Density: }In an ideal case, all the nodes in the graph are connected with each other. That is if there are $N$ nodes in the network then there will be $N$*($N$-1)/2 edges present in the network. However, it may not be the case in reality. Let $E_R$ represent total edges in the network then edge density is defined as the ratio of total edges being present in the network to the total edges that should be present in an ideal case and can be represented as 2*$E_R$/$N$*($N$-1).

\item \textbf{Clustering coefficient (CC): }Similar to edge density, which is a concept about dyad, CC measures the density in terms of triad. The local CC is measured with respect to each node and its neighbors. Local CC is defined as the ratio of the number of edges being present among the neighbors of the node under observation to the total number of edges possible among the neighbors. Average CC is calculated by averaging the CC of all the nodes of the network. Edge Density and CC can be used for inferring how tightly knit a network is.

\item \textbf{Disconnected components: }A connected component is a collection of nodes where every node has a path to every other node in the network. In an ideal case, all the nodes of a network can be reached through some path. However, it is possible that there might not exist a path between every two pairs of nodes in the network. In other words, some  set of nodes might be disconnected with the other set of nodes and thus create disconnected components in the network. 

\end{itemize}

\textbf{Community Detection: }
In a complete connected graph, some of the nodes might be more densely connected with each other compared to rest of the nodes in the network. The set of densely connected nodes forms a community. The concept of community in a graph is analogous to a set of people in a city which are more closely connected with each other due to cultural and religious background compared to other inhabitants of a city. 
In this work, we used Louvain algorithm to detect communities in the network \cite{louvain}. This greedy optimization method is a widely used algorithm because of its fast execution time in handling very large graphs.

%% file: relatedwork.tex
\section{Related Work}
\label{sec:related}
\def\abbre#1{\expandafter\abbreA#1 \relax/ }
\def\abbreA#1#2 {\ifx#1\relax \else\ifnum\uccode`#1=`#1#1\fi\expandafter\abbreA\fi}

\textbf{Social Networks:} Researchers have investigated different social entities in various domains to determine if the concept of flocking behavior is prevalent or not. For example, in politics, researchers investigated the interactions among politicians and citizens on Twitter, and found compelling evidence of homophily \cite{Anders2013}. Another longitudinal study on scientific collaborations also shows strong flocking behavior\cite{Tang2013}. In a different study, researchers investigated ethnic and cultural roles in the creation of friendship among adolescents \cite{Hamm2000}. Surprisingly, the flocking nature is also observed among patients looking for medical treatments \cite{Vieira2013}, as well as among spammers, who end up spamming the same users based on their common intentions \cite{Yehonatan2015}.
Migration behavior due to homophily has also been explored in various domains. In particular mainly with respect to immigrants \cite{ImmigrantBlog}, researchers have studied health among minority members \cite{immigPaper} and the destination decisions made by flocks for migration consider socio-economic parameters  as one of the important factors \cite{Pappalardo2015}. Lu et al. studies the mobile data to understand the migration patterns of a community after a natural calamity \cite{Lu11576} and found that after the earthquake, people migrate to the places where they were making frequent calls before the natural disaster.

\noindent\textbf{Software Engineering:} \abbre{Git Hub} and \abbre{Stack Overflow} have been two popular online peer production sites for programmers for past ten years. Researchers have concluded that the interactions on \abbre{Git Hub} can be viewed as social activities \cite{Thung2013}, and developers' behavior is largely influenced by the awareness of the fact that they are being observed by their peers \cite{Dabbish:2012}. \abbre{Stack Overflow}, a question answering platform predominantly for programming, questions are answered in a median time of 11 minutes, providing quick solutions to technical problems \cite{Mamykina:2011}. Research has explored how \abbre{Stack Overflow} encourages participants to ask "good" questions and to give "good" answers through reputation incentives, such as points and badges  \cite{Capiluppi2013}. 
Past research has studied community formation, or flocks, extensively on \abbre{Git Hub}. Dabbish et al. \cite{Dabbish:2012} highlights many of the motivations behind users forming communities, including the facilitation of communication, streamlining of technical goals, collective inferrence of project outcomes, advancment of technical skills, and managing of reputation. Lima et al. \cite{Lima2014} discussed social interactions on \abbre{Git Hub} and found that active users may not necessarily have a large follower base and the users in close proximity according to geographic location are more ready to interact with each other. Thung et al. \cite{Thung2013} also uncovered and analyzed inter-project and inter-developer relations on \abbre{Git Hub}. Majumder et al. \cite{Majumder2012} researched \abbre{Git Hub} to discover optimal team-formation techniques and algorithms. Yu et al. \cite{Yu:2014} looked at the power of social programming, linking programming social networks to attracting external developers and causing explosive growth in development.
Tsay et al. \cite{Tsay:2014} saw how social connections and interactions influenced accepted pull requests and ultimately the development path of a project. Our paper built upon similar underlying concepts of social interaction on \abbre{Git Hub} but extended them to and focused on the homophily flocking and migration of advocates. 

The only research focused on flocking and migration within \abbre{Stack Overflow} is from Ford et al. \cite{Ford2017a, Ford2017b}. They conducted  a thorough research on peer parity with women in \abbre{Stack Overflow}. They analyzed how women interacted, formed communities, and helped each other within this male-dominated field.Researchers have also performed cross-site studies of users within \abbre{Git Hub} and \abbre{Stack Overflow}. Vasilescu et al.\cite{Vasilescu2013} discovered that a user's activity and participation on \abbre{Stack Overflow} correlates with their coding activity on \abbre{Git Hub}. Badashian et al.\cite{Badashian:2014} performed an in depth analysis and followed inter-network activity over a five-year period to look for patterns between activity on the two websites. However, their results showed that there existed little correlation. Our work is different from past research as we study the advocates - developers who are active in both \abbre{Git Hub} and \abbre{Stack Overflow}, using social network analysis to understand the flocking and migration patterns of advocates across both the sites.

%% file: methodology.tex
\section{Methodology}
\label{sec:method}
\def\abbre#1{\expandafter\abbreA#1 \relax/ }
\def\abbreA#1#2 {\ifx#1\relax \else\ifnum\uccode`#1=`#1#1\fi\expandafter\abbreA\fi}

We investigated the flocking and migration patterns of advocates in and across peer production sites using network science theories. 

\subsection{Dataset}
\label{subsec:dataset}

We collected \abbre{Git Hub} related data using GHTorrent - a public off-line mirror of \abbre{Git Hub} data offered using the \abbre{Git Hub} REST API. 
We extracted and processed the data for our research questions. 
We Used the `commits' and `projects' features to filtered out the \texttt{user\_id} and \texttt{commit\_SHA} for each advocate in our list. 

\noindent\textbf{\abbre{Git Hub}:}
To collect the data related to advocates making changes to a file in a project, we web-crawled \abbre{Git Hub} with \texttt{Python} using the package \texttt{scrapy}. As GHTorrent did not have data on who edited what file, this was done by generating the links by appending the \texttt{commit\_SHA} to the url on \abbre{Git Hub}'s search page and following the web page of that commit. From the commit page, we could pull the list of all files that were affected with that commit. Also note that we worked with the GHTorrent data dump of Sept'16.

\noindent\textbf{\abbre{Stack Overflow}:}
To collect the \abbre{Stack Overflow} related data we have used BigQuery dataset \cite{sovdata}, which gets updated on a quarterly basis. This dataset includes an archive of \abbre{Stack Overflow} content, which includes \texttt{posts, votes, tags, answers, comments, and badges}. 

\noindent\textbf{Common Advocates between two sites:}
To find the common advocates on both sites, we selected the advocates who provided \abbre{Git Hub} links in their \abbre{Stack Overflow} profiles. We retrieved 12,578 common advocates between the two sites using this technique. Unlike past studies \cite{Lee2017}, we could not use emails to identify common advocates across \abbre{Git Hub} and \abbre{Stack Overflow} because \abbre{Stack Overflow} no longer provides email ids of advocates in their public database to protect user privacy.

\subsection{Model}
\label{subsec:model}

We used graph theory to analyzed the network of \abbre{Git Hub} and \abbre{Stack Overflow} advocates. Let $G (N,E,P,F)$ be an undirected graph representing the advocates and their connections in a platform, where $N$ represents all the advocates of a production site under investigation, and E represents a set of connections among the advocates.  $P$ represents a set of all the projects in the case of \abbre{Git Hub} and posts for \abbre{Stack Overflow}. $F$ represents a set of all the files for \abbre{Git Hub} and all the interactions for \abbre{Stack Overflow}. Two advocates, $N_i$ and $N_j$, in $G$, are connected if they have committed to the same file $F$ in a \abbre{Git Hub} project or interacted by answering/commenting to a post in \abbre{Stack Overflow}.

\subsection{Entities}

Based on this model, the basic entities for \abbre{Git Hub} and \abbre{Stack Overflow} are:

\noindent\textbf{\abbre{Git Hub}:} \noindent\texttt{Project:} Each \abbre{Git Hub} project repository represents a separate community network, where one or more advocates might be contributing to the project. Advocates can commit, fork, or pull-request for each project.

\noindent\texttt{Files:} A project consist of multiple files. Advocates can add, delete or modify files collaboratively or individually.

\noindent\textbf{\abbre{Stack Overflow}:} \noindent\texttt{Post:} Each \abbre{Stack Overflow} post represents a separate community network of one or more advocates. Advocates generally ask, answer, or comment on the posts.

\noindent\texttt{Interactions:} An interaction can be all the ways a advocates may interact with others advocates like to ask questions, provide answers,  comment, mark a post favorite, or cast votes.

\subsection{Detecting Communities}

Based on the list of common advocates on \abbre{Git Hub} and \abbre{Stack Overflow}, we detected the communities of these advocates within and across \abbre{Git Hub} and \abbre{Stack Overflow} as follows:

\noindent\textbf{Approach for \abbre{Git Hub}:} We selected `editing a common file' to define interactions, as configuration management systems allow advocates to check out files for making changes, which facilitates rapid parallel development. Hence, two advocates working on either the same file or interrelated files need to communicate with each other to avoid direct or indirect conflicts \cite{Sarma:2007}.
With the list of all files collected from the web crawler, written in python we then created an edge between two advocates if they have worked on a common project.  After determining the pairs of advocates who contributed to the same file in each project, we collected 860 pairs of advocates across all the project files. Some of the advocates from the list were removed, as either they did not contributed or projects were not accessible for public. 


\noindent\textbf{Approach for \abbre{Stack Overflow}:} 
To collect the dataset, we used \textit{posts\_questions}, \textit{posts\_answers} and \textit{comments} tables from \abbre{Stack Overflow} database. We generated a list of 9445 posts, which consisted of all the interactions between advocates from our common list. 
	
Finally, the collected data was organized to indicate all the possible interactions within a post in the form of answers or comments. We also found some people who commented and answered on their own posts, but we filtered those out, as they were not useful in community detection. 



\noindent\textbf{Detecting communities and migration:} In both the datasets, we generate communities using \texttt{Louvain} \cite{louvain} method.
We looked into migration both within and across \abbre{Stack Overflow} and \abbre{Git Hub} by noting flock movement in posts/ projects. We created all possible pairs of nodes among all the community members and for each pair, we checked for a match among the other communities and counted the migrations. Similarly, match cases were also computed between \abbre{Stack Overflow} and \abbre{Git Hub} to extract communities that migrated between the platforms.


\subsection{Characteristics of Advocates}
\label{subsec:char_adv}
We analyzed the following characteristics of advocates.
\begin{multicols}{2}
\noindent\textbf{\abbre{Git Hub}}
\begin{itemize}
\item Language of Expertise
\item Project Owned
\item Location
\end{itemize}

\noindent\textbf{\abbre{Stack Overflow}}
\begin{itemize}
\item Language of Expertise
\item Reputation
\item Location
\end{itemize}
\end{multicols}

\texttt{Language of Expertise} can be categorized as both social/technical skill \cite{Sharma2016} - suggests the advocates' propensity towards the type of programming language the advocate is an expert in \abbre{Git Hub} or \abbre{Stack Overflow}. Another important criterion is \texttt{Projects Owned}, categorized as technical skill and \texttt{Reputation} as social skill \cite{Sharma2016}, which are indicators of advocates expertise on two sites.  We calculated the \texttt{Projects Owned} and \texttt{Reputation} for each community by $abs(diff(R_{i}, R_{i+1},...,R_{n})) \forall i,$ where $i$ is the number of advocates and $R_{i}$ is the \texttt{Reputation} score or \texttt{Projects Owned} of the $i^{th}$ advocate. To consider only strong relationships between advocates, we accounted for only those pairs in which the difference of the \texttt{Reputation} scores of two advocates were less than or equal to the threshold $\theta$. The threshold $\theta$ is calculated according to the following equation: $mean(abs(diff(R_{i},R_{i+1},$ $...,R_{n-1},R_{n})_{j}))$ $\forall i,j$ where $i$ is the number of advocates and $j$ is the number of communities. The notion is, lower the value of \texttt{Projects Owned} or \texttt{Reputation} Score, means closer the advocates are in terms of strata. The \texttt{Location} criteria presents insights about whether advocates prefer to collaborate based on geographic locations - as Lima et al. mentioned users tend to interact with people that are close, as long-range links have a higher cost \cite{Lima2014}. We found the 301 missing data points out of 951 in \abbre{Git Hub} and 166 advocates out of 1104 who did not have \texttt{Location} attribute for \abbre{Stack Overflow}. Since the percentage of missing data is 31\% (for \abbre{Git Hub}), and approximately 15\% (for \abbre{Stack Overflow}), we still reported for this attribute, as past research \cite{Lima2014} found the close geographic proximities leads to more collaborations on \abbre{Git Hub}.

%% file: results.tex
\section{Results}
\label{sec:results}
\def\abbre#1{\expandafter\abbreA#1 \relax/ }
\def\abbreA#1#2 {\ifx#1\relax \else\ifnum\uccode`#1=`#1#1\fi\expandafter\abbreA\fi}
To observe the flocking and migration behavior of the advocates in intra and inter-peer production sites we performed microscopic analysis, in which we performed detailed empirical analysis with respect to three research questions. And we analyzed the advocates' behavior for \abbre{Git Hub} and \abbre{Stack Overflow} with microscopic analysis.


\subsection{Macroscopic Analysis}
\label{subsec:4.1}

We collected all the nodes (advocates) across all the different projects/ posts to understand the bird's-eye view of the \abbre{Git Hub} and \abbre{Stack Overflow} network. Table \ref{tbl:nwMet} provides information about various metrics of these networks.

The total nodes in \abbre{Git Hub} were less than \abbre{Stack Overflow}. The network of \abbre{Git Hub} was more spread out (Diameter and APL) and less dense (Density) compared to \abbre{Stack Overflow}. However, in \abbre{Git Hub}, friends of a friend property is more visible (CC) when compared with \abbre{Stack Overflow}. Among both the networks, \abbre{Stack Overflow} was less disconnected. Based on our macroscopic analysis, we present four observations as follows:


\begin{table}
\centering
\begin{tabular}{|l|l|l|}
\hline
 Metrics                           & \abbre{Git Hub}      & \abbre{Stack Overflow} \\ \hline
\# Nodes                          & 951         & 1104          \\ \hline
\# Edges                          & 793         & 1173          \\ \hline
Average Degree                  & 0.8338         & 1.0625            \\ \hline
Average Path Length               & 7.445656    & 6.353188      \\ \hline
Diameter                          & 18          & 16            \\ \hline
Density                           & 0.001755493 & 0.001926564   \\ \hline
Clustering coefficient            & 0.279148    & 0.0227758     \\ \hline
\# disconnected components       & 281         & 123           \\ \hline
\end{tabular}
\caption{Macroscopic Analysis}
\label{tbl:nwMet}
\end{table}

\begin{figure}
\hfill
\subfigure[\abbre{Stack Overflow}]{\includegraphics[width = 5cm]{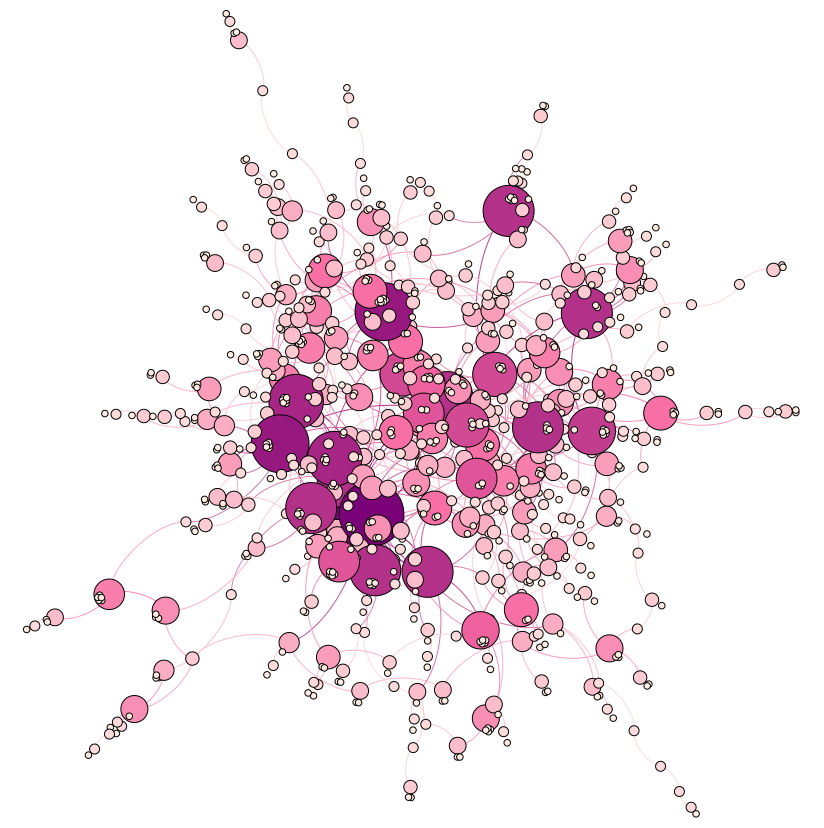}}
\hfill
\subfigure[Git Hub]{\includegraphics[width = 5cm]{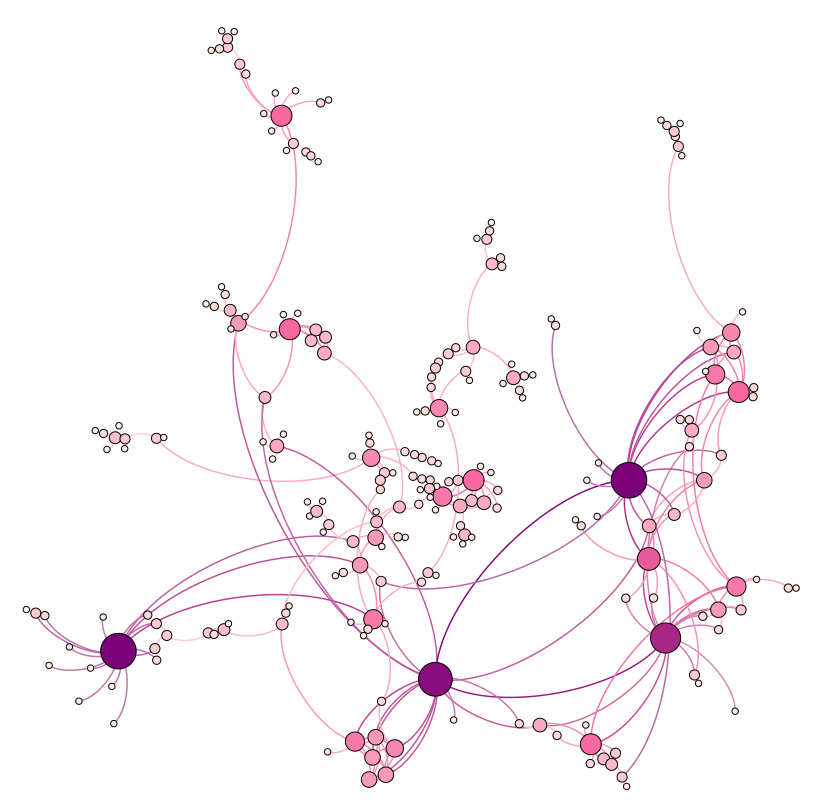}}
\hfill
\caption{Biggest connected components}
\label{BigConnCompGit}
\end{figure}

\begin{itemize}

\item \textbf{Observation 1 -- Inter-component interactions:  }In \abbre{Git Hub}) and \abbre{Stack Overflow} networks there is a one big connected component consisting of 235 (~24.7\%) and 821 (74.3\%) nodes for \abbre{Git Hub} and \abbre{Stack Overflow} respectively (Figures not shown due to space limitation). Also, the number of disconnected components in \abbre{Git Hub} (281) is much more than \abbre{Stack Overflow} (123) compared to the total nodes present in each of the networks. This indicates that advocates in \abbre{Git Hub} are more isolated in general. In addition, low average degree in \abbre{Git Hub} also supports this assumption, which basically means that on an average a node interacts with just one other node.
Figure \ref{allnodes_\abbre{Git Hub}}(a) (for \abbre{Git Hub}) and Figure \ref{allnodes_\abbre{Git Hub}}(b)(for \abbre{Stack Overflow}) gives a sense of the disconnectedness in both of these networks.
\item \textbf{Observation 2 -- Intra-component interactions: }The high value of Clustering coefficient for \abbre{Git Hub} indicates that within a disconnected component the advocates work closely together. This is in contrast to inter-component interaction, which appears weak.

\item \textbf{Observation 3 -- Interaction Patterns:} Figures for the biggest connected components for \abbre{Git Hub} (Figure \ref{BigConnCompGit}(a)) and \abbre{Stack Overflow} (Figure \ref{BigConnCompGit}(b)) show that in \abbre{Git Hub} there are very few nodes who have high degree of interactions (node size is proportional to the degree/interactions) and most of the nodes work with a small number of advocates. In other words, very few developers tend to work in larger teams and most of the developers tend to work in small teams. However, this is different from \abbre{Stack Overflow}, where one can observe various advocates having varying sizes of interactions.

\begin{figure}
\hfill
\subfigure[\abbre{Stack Overflow}]{\includegraphics[width = 5cm]{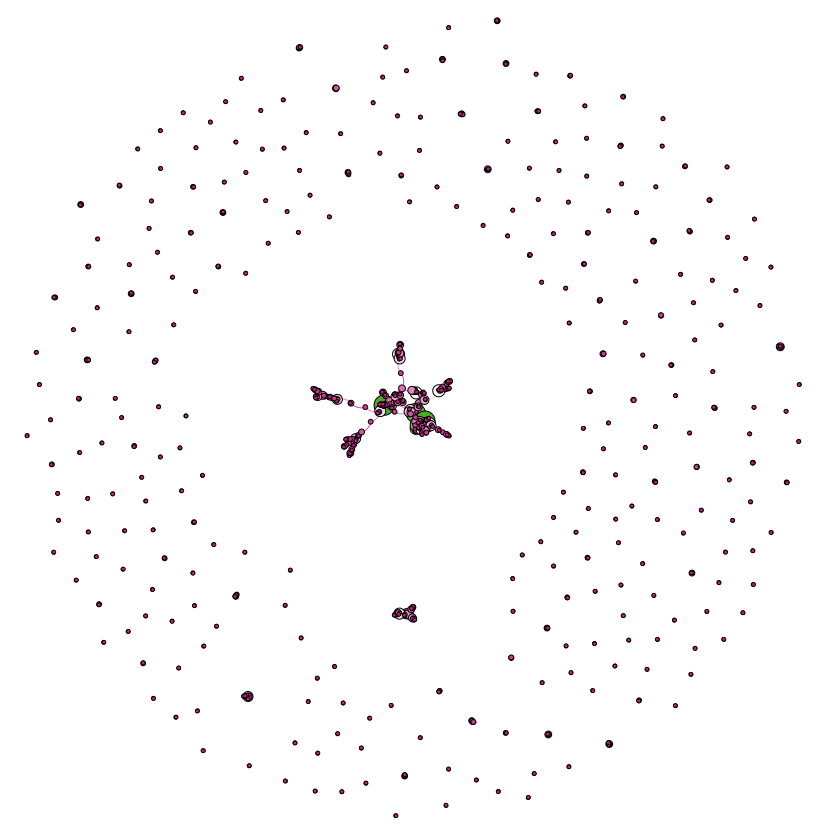}}
\hfill
\subfigure[Git Hub]{\includegraphics[width = 5cm]{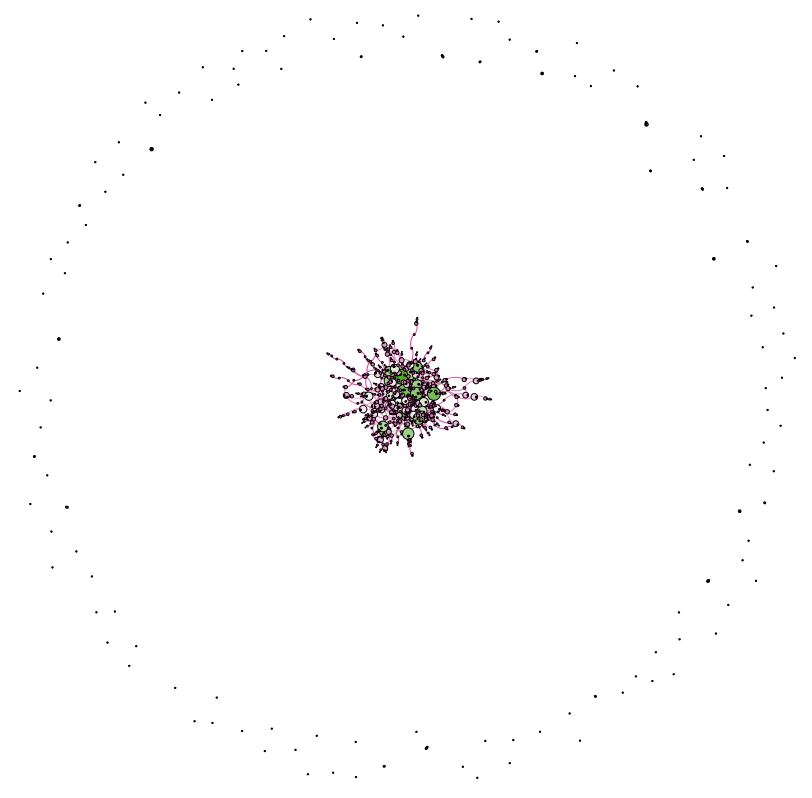}}
\hfill
\caption{All the nodes (including disconnected components)}
\label{allnodes_\abbre{Git Hub}}
\end{figure}

\item \textbf{Observation 4 -- Reaching out to the fellow advocates:  }Relative high values of diameter and average path length in \abbre{Git Hub} indicate that spreading a message to fellow advocates will take more time in \abbre{Git Hub} than in \abbre{Stack Overflow}.
\end{itemize}

\begin{tcolorbox}
We found that \abbre{Git Hub} is not only more disconnected but also has smaller degree of interactions compared to \abbre{Stack Overflow}.
\end{tcolorbox}

\subsection{Microscopic Analysis}
\label{subsec:4.2}

As in the Macroscopic analysis, we cannot
To observe the flocking and migration behavior of the advocates in intra and inter-peer production sites we performed microscopic analysis, in which we performed detailed empirical analysis with respect to three research questions.

\begin{table}
\begin{center}
\begin{tabular}{|c||l|l|l|l|l|l|l|l|l||l|l|}
  \hline
  \multirow{1}{*}{Number of} 
      & \multicolumn{9}{c||}{Flocks} 
          & \multicolumn{2}{c|}{Migration}                      \\  \cline{1-12}
 Flocks involved & 1 & 2 & 3 & 4 & 5 & 6 & 7& 8 &9 &1 &2		\\  \hline
Advocates involved& 852 & 132 & 37 & 9 & 7 & 2 & 2& 1& 1& 165 &6\\  \hline
\end{tabular}
\end{center}
\caption{Migration of advocates across flocks with-in \abbre{Git Hub}.}
\label{rep}

\end{table}

\subsection{RQ1: Do advocates flock together in a peer production site?}
\label{subsubsec:4.2.1}

To understand the flocking behavior of the advocates, we identified flocks using \texttt{Louvain} algorithm.

\noindent\textbf{(a) Flocks on \abbre{Git Hub}:} We found that approximately 7.5\% (951/12,578) of advocates flocked across \abbre{Git Hub}. In total, 780 flocks were formed by these advocates, with sizes ranging from 2 to 11 advocates in each flock (Refer to Table \ref{git1}).

\begin{table}
\begin{center}
    \begin{tabular}{|l|c|c|c|c|c|c|c|c|c|}
        \hline
            \multirow{1}{*}{ } 
                & \multicolumn{7}{c|}{\abbre{Git Hub}}
                    & \multicolumn{2}{c|}{StackOverflow}           \\ \cline{1-10}
                     \# Advocates					& 2 & 3 & 4 & 5 & 6 & 7 & 11 & 2 & 3	    \\ \hline
                     \# of Flocks Found  	        & 707 & 57 & 8 & 5 & 1 & 1 & 1 	& 1229 & 21 \\ \hline
                     \# of Flocks Migrated          & 109 & 3 & 0 & 0 & 0 & 0 & 0 & 57 & 0      \\ \hline
                      \# of times Flocks Migrated   & 609 & 3 & 0 & 0 & 0 & 0 & 0 & 125 & 0	    \\ \hline

    \end{tabular}
\end{center}
\caption{Overall Flocking and Migration pattern for advocates}
\label{git1}
\end{table}

\begin{table}
\begin{center}
\begin{tabular}{|c||l|l|l|l|l|l|l|l|l|c|l|l|c||l|c|}
  \hline
  \multirow{1}{*}{Number of} 
      & \multicolumn{13}{c||}{Flocks} 
          & \multicolumn{2}{c|}{Migration} \\   \cline{1-16}
Comm. involved 	&1 &2 &3 &4 &5 &6 &7 &8 &9 &10/11/12 &13 &14 &15 or More &1 &2 or More\\ \hline
Adv. involved 	&662 &172 &92 &59 &27 &19 &13 &15 &6 &5 &4 &2 &13 &0 &256 		 \\ \hline
\end{tabular}
\end{center}
\caption{Migration of advocates across flocks with-in \abbre{Stack Overflow}.}
\label{rep2}
\end{table}


\noindent\textbf{\textit{How do the advocates tend to flock?}} As we tried to extrapolate a general pattern behind the advocates forming flocks, we observed that 951 advocates formed different sizes of flocks. Table \ref{rep} displays the advocates found in different flocks within \abbre{Git Hub}. To add, 852 advocates were found that belonged to 1 flock, 132 advocates were part of 2 flocks and just 1 advocate belonged to 9 flocks.  
Thus, it's quite evident from Table \ref{rep} that the general proclivity of advocates is to disperse into congenial groups and form single flocks among themselves than joining multiple flocks.

\begin{table}
\begin{center}
\begin{tabular}{|l|c|c|}
  \hline
  \multirow{1}{*}{ }
     	&\multicolumn{1}{c|}{\abbre{Git Hub} Flocks (Found/Total)}
     	    &\multicolumn{1}{c|}{StackOver Flow Flocks (Found/Total)} \\ \hline
 Language of Expertise 	&780/780 	&1024/1250						\\ \hline
 Project Owned  		&557/780 		&NA					        \\ \hline
 Location* 				&356/601 	&354/932	 					\\ \hline
Reputation                &NA          &1057/1250                   \\  \hline
\end{tabular}
\end{center}
\label{char_flock}
\caption{Characteristics that led to Flocking}
\label{char_git}
\end{table}

 

\noindent\textbf{\textit{What characteristics motivated advocates to flock?}} We next analyzed all 780 flocks to understand how the characteristics of the advocates played a role in forming the flocks.
 Table \ref{char_git} shows the characteristics of flocks. All 780 flocks' respective advocates have one or more common \emph{programming language} of Expertise. For the Projects Owned characteristic, we found only 71.4\% flocks were below the threshold $\theta$ (mentioned in Section 2.5).
Finally, we found 356 (59\%) flocks had advocates belonging either from the same \textit{country} or \textit{continent}. (Note: Only 601 flocks contained location information.) Among these flocks, 96 had advocates from the same \textit{continent} and 260 from the same \textit{country}. These results suggest that advocates having same field of interests form communities (birds of a feather flock together). Hence, people knowing same languages like minded (owners of projects) and living in close proximity tend to create flocks.

\noindent\textbf{(b) Flocks on \abbre{Stack Overflow}:} In the case of \abbre{Stack Overflow}, we found 1104 advocates (out of 12.5K) that flocked in different sizes, which is approximately 8.7\%, as evidenced by Table \ref{git1}. 
These 1104 advocates were observed to form 1250 flocks of sizes 2 and 3. From Table \ref{git1} we can see 1229 flocks were formed between 2 advocates and only 21 flocks were formed between 3 advocates.

\noindent\textbf{\textit{How do the advocates tend to flock?}} 
We found 662 advocates who were part of 1 flock, 172 advocates who were a part of 2 flocks, 5 advocates who were involved in 11 flocks, and so on (Table \ref{rep2}). Even in the case of \abbre{Stack Overflow}, the general trend in forming flocks is similar to \abbre{Git Hub}. 

\noindent\textbf{\textit{What characteristics motivated advocates to flock?}} We analyzed the characteristics that led the flocks to migrate over different posts. Table \ref{char_migrate} details the characteristics of advocates in a flock. We used tags from \textit{posts} to extract out topics on which a particular advocate preferred to answer or ask a question. We had 1250 flocks of varied sizes, out of which we found the 1024 (81.92\%) flocks that stayed together due to one or more interest in topic. For the Reputation criteria, we found 1057 (84.5\%) of such flocks that were below the threshold $\theta$. For location characteristics, we had to narrow down our search to 932 (74.56\%) flocks out of 1,250, as some advocates didn't have the location data. We qualitatively analyzed each flock and our match criteria were either \textit{continent} or \textit{country}; we found 221 flocks that had advocates from same continent and 133 from same country. In total, 37.9\% flocks were formed either based on matching country or continent. Similar to \abbre{Git Hub}, we found that same interest advocates create flocks in \abbre{Stack Overflow}. 


\begin{tcolorbox}
We found around 8\% advocates form flocks on \abbre{Git Hub} and 9\% on \abbre{Stack Overflow}. Flocks of two were most common. Further, two advocates who had same field of interest flocked together on \abbre{Git Hub} and \abbre{Stack Overflow}. 
\end{tcolorbox}

\subsection{RQ2: How do flocks of advocates migrate within a peer production site?}

We investigated the migration pattern of flocks between \abbre{Git Hub} and \abbre{Stack Overflow}, i.e., flock of advocates moving from one project/post to another. 

\begin{figure}
    \centering
    \includegraphics[width=.8\textwidth]{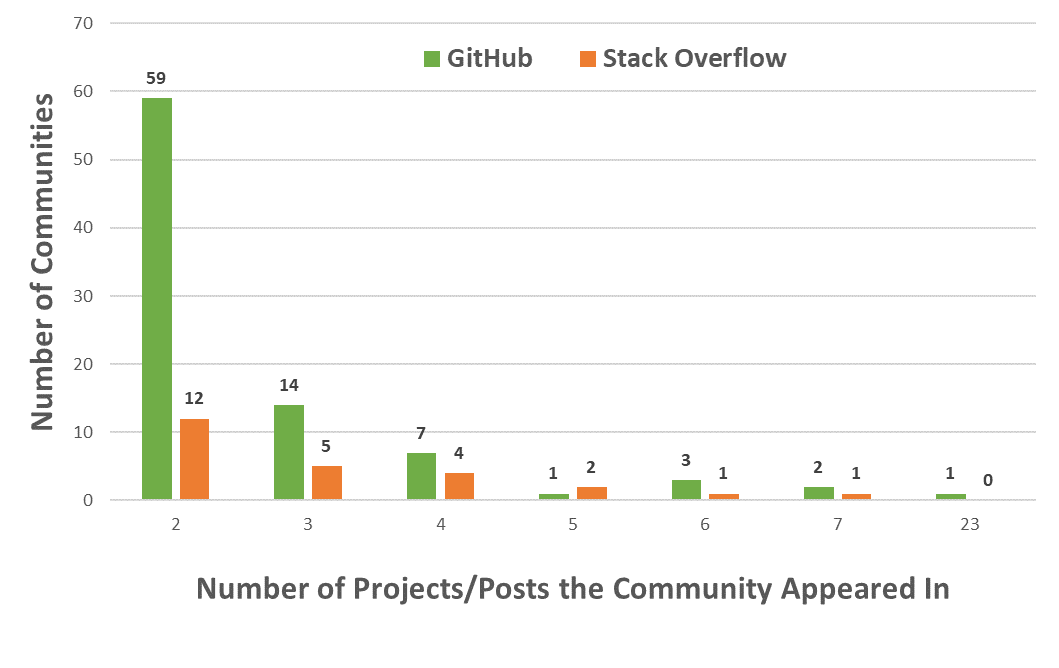}
    \caption{\# of Flocks vs \# of Projects Appeared in GitHub}
    \label{sov_comvsproj}
\end{figure}

\noindent\textbf{(a) Migration on \abbre{Git Hub}} We wanted to understand the migration patterns of advocates on GithHub.  Table \ref{char_migrate} summarizes the number of flocks migrated and number of times these flocks migrated across different projects. We found that 109 flocks that had two advocates migrated about 609 times over different projects and 3 flocks that had 3 advocates that migrated 3 times. To understand the diversity of projects advocates flocked together, we investigated the total number of flocks that appeared in multiple projects. Figure \ref{sov_comvsproj} summarizes  as can be seen 59 flocks contributed to 2 projects. Interestingly, all the projects had more than one flock in it.

\noindent\textbf{\textit{How do the advocates migrate on a code hosting site?}} As mentioned earlier, we found 951 advocates that formed different sizes of flocks on \abbre{Git Hub}, and only a total of 171 advocates migrated across the different flocks. Table \ref{rep} shows that 165 advocates were found in at least 165 single unique flock and only 6 advocates were found in at least 2 different flocks. 

\begin{table}
\begin{center}
\vspace{-10pt}
\begin{tabular}{|l|c|c|}
  \hline
  \multirow{1}{*}{ } 
       & \multicolumn{1}{c}{\abbre{Git Hub} Flocks (Found/Total)}
            & \multicolumn{1}{|c|}{StackOverflow Flock(Found/Total)} \\ \cline{1-3}
 
Language of Expertise 		&112/112	&51/57 					\\ \hline
Project Owned				&77/112		&NA				        \\ \hline
Location*	  				&54/88 		&24/44					\\ \hline
Reputation                  &NA         &38/57                  \\ \hline
\end{tabular}
\end{center}
\caption{Characteristics that led to migrate}
\vspace{-30pt}
\label{char_migrate}
\end{table}

 
\noindent\textbf{\textit{What characteristics motivate a flock of advocates to migrate?}} Table \ref{char_migrate} summarizes the characteristics of flocks of advocates. For Language of Expertise, all 112 flocks had advocates contributes on code with the same programming language. As for the Project Owned, we found 68.75\% flocks that had difference lower than $\theta$. We found 61.36\% flocks had advocates who belonged to either the same country or continent.  Thus it can be concluded that advocates belonging to the same field of expertise migrate together.

\noindent\textbf{(b) Migration on \abbre{Stack Overflow}} We observed 57 flocks of two advocates that migrated 125 times across different posts. However, no flock of 3 advocates migrated (Refer Table \ref{char_migrate}). Figure \ref{sov_comvsproj} shows the flocks that appeared across different posts. We recorded 12 flocks that showed up in 2 posts, 5 flocks in 3 posts and just 1 flocks that appeared in 7 posts. These results indicate that flocks were not active in different posts on \abbre{Stack Overflow}.

\noindent\textbf{\textit{How does advocates migrate on a Q/A site?}} In this platform, 1,104 advocates formed different sizes of flocks; however, only 256 of them migrated. In Table \ref{rep2} under migration column, we can see that no advocate who was part of a single flock migrated. However, advocates who were involved in more than 1 flock migrated. This finding is in direct contrast with the advocates migration pattern which we observed in \abbre{Git Hub}. Thus in \abbre{Stack Overflow} advocates do like to form flocks with like-minded people, but when it comes to migration, they might not keep those friendships over time. 

\noindent\textbf{\textit{What characteristics motivate a flock of advocates to migrate?}} We extracted out characteristics to understand the motivations behind migration of flocks. Table \ref{char_migrate} summarizes the three characteristics of flocks when moving between posts. For Language of Expertise, we found 89.4\% of the flocks that migrated had advocates who worked on the same language as depicted in. For Reputation,  66.66\% flocks that had a difference in reputation below the threshold migrated. In the case of Location, we found 13 missing data points, and found 54.5\% of flocks who have advocates from same the continent or country tend to stick together as they migrate. Hence, the migration in \abbre{Stack Overflow} can be attributed among advocates with same language and reputation but geographically distributed.


\begin{tcolorbox}
Our results indicate that on \abbre{Git Hub}  two advocates who flock based on the same interest migrate across projects while on \abbre{Stack Overflow}, a pair of advocate might break their bonds. Further, while migrt. the \abbre{Stack Overflow} adv. may learn/help beyond geographic locations.
\end{tcolorbox}

\subsection{RQ3: Do the flocks of advocates migrate beyond a single peer production site?}
We wanted to see if the flocks of advocates migrated across code hosting and Q/A platforms. We found only 3 flocks out of 1250 (\abbre{Stack Overflow})/ 780 (\abbre{Git Hub}) migrated across the sites. In terms of advocates, only 6 out of 951 advocates in \abbre{Git Hub} and 1104 advocates in \abbre{Stack Overflow} migrated across the sites.

As we tried to observe how these 6 advocates migrated with-in the sites, we found these 6 advocates paired among themselves and formed 3 flocks and these flocks never migrated both in \abbre{Git Hub} and \abbre{Stack Overflow}. Also note, that 2 out of the 3 flocks were a subset of a larger flock (not necessarily the same) consisting of 3 advocates. 

\noindent\textbf{\textit{What characteristics motivate flocks to migrate across platforms?}} We also extracted out the characteristics of the advocates that migrated between \abbre{Stack Overflow} and \abbre{Git Hub}. For the first flock, both the advocates belong from the same country and both had a reputation of  34,900. However, both these advocates had just one field of interest (Language of Expertise) in common. However, the second flock had a strong relation between their field of interest matching up-to 3 tags and difference in reputation score 43,863. However, their location was vastly different, each belonging to a different continent. Finally, for the third flock, both advocates belonged from the same continent, but had a difference in reputation score of 1524 with only one field of interest in common. Only three flocks migrated across peer production sites, among all three flocks as they had one or more common characteristics.

\begin{tcolorbox}
Only three flocks migrated across peer production sites, and they had one or more common characteristics.
\end{tcolorbox}

%% file: disscon.tex
\section{Conclusion and Implication}
\label{sec:concl}
\def\abbre#1{\expandafter\abbreA#1 \relax/ }
\def\abbreA#1#2 {\ifx#1\relax \else\ifnum\uccode`#1=`#1#1\fi\expandafter\abbreA\fi}

In this paper, we analyzed the flocking-migration behavior of 12.5K advocates who were active on \abbre{Git Hub} and \abbre{Stack Overflow}. 
Our results the analysis verify that advocates do flock and migrate. We found that 7.5\% of the advocates create flocks on \abbre{Git Hub} and 8.7\% on \abbre{Stack Overflow}. Further, these flocks of advocates migrate on an average of 5 times on \abbre{Git Hub} and twice on \abbre{Stack Overflow}. Our results show a general trend that advocates in \abbre{Git Hub} tend to work in small teams. This pattern may induce less flocking and migration behavior among the advocates. Further, advocates in \abbre{Git Hub} are bound by long-term project interactions, which may lead them to be more selective of their collaborators. Unlike \abbre{Git Hub}, \abbre{Stack Overflow}'s interactions are sporadic and short-term, resulting in the creation of more connections. Our findings open new opportunities for software practitioners, researchers, and tool builders to study and support the flocking-migration of advocates in and across different peer-production sites.

Our findings  have a number of implications for tool builders to facilitate flocking and migration behavior within and across peer production sites.

\noindent\textbf{Searching Code based on Social Interactions:} The understanding of the flock formation can help in the design of code-searching tools based on their social interactions. This new paradigm can help in searching for trusted code examples in one's own social network. For example, these tools could leverage socio-technical skills from peer production sites to advertise, monitor, and assess the quality of code contributions using psycho-physiological measures to evaluate task difficulty.

\noindent\textbf{Migration Within a code hosting site:} Our results suggest that flocks formed in \abbre{Git Hub} tend to be small, but are restricted more to coordination among individuals and technology specific to their projects. Hence, currently these flocks exist in isolation. Such behavior was also observed by Datta et al.  \cite{Datta2015}. Their results suggested that developers work in isolation, which may have been facilitated by the distinct dynamic of code peer reviews. Hence, we need to study ways to motivate and design tools to support migration within a code hosting site. 

\noindent\textbf{Migration across peer production site:} Results suggests little migration of flocks across the peer production sites. Hence, there is a need to build predictive analytics and recommendation applications for supporting migration across peer production sites. For instance, prediction based models that observe the socio-technical activities of advocates on multiple peer production sites could recommend advocates for collaboration. The models could recommend flocks across communities on both code hosting and Q\&A sites.

\noindent\textbf{Automated-task generation for learners:}  Based on the communication and incentive structures of peer production sites, we can develop an automated approach where small tasks can be selected from the ecosystem of projects to help newcomers learn new skills that match their career goals.

%% file: threat.tex
\section{Threats to validity}
\label{sec:6}
Our research findings may be subject to concerns, but we have taken steps to eliminate the impact of these possible threats and discuss them below.
    
\textit{Bias due to sampling and dataset:} We collected advocates from a single project-hosting website, GitHub, and a single question and answer website, Stack Overflow. Thus our conclusions may not be perfectly generalizable. However, these are the most popular peer production sites, and they represent the vast majority of developers for creating open source softwares.

Secondly, our dataset of advocates is 12.5K, which is much smaller than the past research \cite{Badashian:2014}, as they collected 92K common developers using MD5 hashes, which are no longer accessible due to privacy reasons. Hence, our data is not completely inclusive, and the source of data may have evolved. However, we argue that although our data is small, it is more accurate, as we used GitHub links from the Stack Overflow profiles to collect advocates. 
    
Thirdly, a threat may arise from the source of data, GHTorrent, as a past analysis has indeed shown that all of GitHub may not be fully duplicable through the REST API \cite{G13}. However, GHTorrent has been used extensively by researchers \cite{Lee2017, G13,Badashian:2014} and does include a valid dataset of GitHub. 

\textit{Bias due to used metrics:} There can be a threat based on the way we defined communities. i.e., advocates who contributed to the same file in a project. Another threat can exist as we have not considered time while forming communities on GitHub. Two developers who worked on the same file years apart probably should not be considered a community. We argue that this scenario can not occur in our analysis as we filtered out any communities that only appear for one file. Moreover, our metric is only one of the many possible ways to define and detect communities. The data collected from BigQuery for Stack Overflow returned data tuples with comments associated with posts, which are not representative of real scenarios, where comments are associated with either an answer or a question of a post. Hence, our community construction may not be valid for interaction of associates when comments are included, but we argue that it does represent an interaction in a post.

Further, we have selected few characteristics for GitHub and Stack Overflow to do in-depth analysis. Other characteristics such as ages, up-votes, down-votes, number of files modified, type of project, etc. were not considered. Considering these may give other potential insights but still our results are valid for the three characteristics we selected for our dataset.

\textit{Bias due to community detection algorithm:} The selection of Louvain community detection algorithm is based on the fact that it is one of the most popular and efficient algorithms. However, it only returns communities which are either very small or large. Thus, this may affect the community analysis in general. However, all other  community detection algorithms have their own limitations.

%% file: conclusion.tex
\section{Conclusion}
\label{sec:concl}
\def\abbre#1{\expandafter\abbreA#1 \relax/ }
\def\abbreA#1#2 {\ifx#1\relax \else\ifnum\uccode`#1=`#1#1\fi\expandafter\abbreA\fi}

In this paper, we analyzed the flocking and migration behavior of 12.5K advocates-developers who were active on \abbre{Git Hub} and \abbre{Stack Overflow}. 
Our results from macroscopic and microscopic analysis verify that advocates do flock and migrate. We found that 7.5\% of the advocates create flocks on \abbre{Git Hub} and 8.7\% on \abbre{Stack Overflow}. Further, these flocks of advocates migrate on an average of 5 times on \abbre{Git Hub} and 2 times on \abbre{Stack Overflow}. Our results show a general trend that advocates in \abbre{Git Hub} tend  to work in small teams. This pattern may induce less flocking and migration behavior among the advocates. Further, advocates in \abbre{Git Hub} are bound by long-term project interactions, which may lead them to be more selective of their collaborators. Unlike \abbre{Git Hub}, \abbre{Stack Overflow}'s interactions are sporadic and short-term, resulting in the creation of more connections. 

Our findings open new opportunities for software practitioners, researchers, and tool builders to study and support the flocking and migration of advocates in and across different peer production sites.

%% file: main.bbl
\begin{thebibliography}{10}
\providecommand{\url}[1]{{#1}}
\providecommand{\urlprefix}{URL }
\expandafter\ifx\csname urlstyle\endcsname\relax
  \providecommand{\doi}[1]{DOI~\discretionary{}{}{}#1}\else
  \providecommand{\doi}{DOI~\discretionary{}{}{}\begingroup
  \urlstyle{rm}\Url}\fi

\bibitem{githubdata}
Ghtorrent db (2017).
\newblock
  \urlprefix\url{https://bigquery.cloud.google.com/dataset/ghtorrent-bq:ght}.
\newblock Accessed: 2018-04-10

\bibitem{sovdata}
Bigquery stack overflow db (2018).
\newblock
  \urlprefix\url{https://bigquery.cloud.google.com/table/bigquery-public-data:stackoverflow}.
\newblock Accessed: 2018-05-10

\bibitem{Badashian:2014}
Badashian, A.S., Esteki, A., Gholipour, A., Hindle, A., Stroulia, E.:
  Involvement, contribution and influence in github and stack overflow.
\newblock In: CSCW, pp. 19--33. IBM Corp. (2014)

\bibitem{louvain}
Blondel, V.D., Guillaume, J.L., Lambiotte, R., Lefebvre, E.: Fast unfolding of
  communities in large networks (2008)

\bibitem{Capiluppi2013}
Capiluppi, A., Serebrenik, A., Singer, L.: Assessing technical candidates on
  the social web.
\newblock IEEE Software \textbf{30}(1), 45--51 (2013)

\bibitem{Yehonatan2015}
Cohen, Y., Hendler, D.: Birds of a feather flock together: The accidental
  communities of spammers.
\newblock In: IEEE/ACM ASONAM, pp. 986--993 (2015)

\bibitem{Vieira2013}
D., F.N., Pascale, Q.: Birds of a feather flock together ... definition, role
  and measure of congruence: An application to sponsorship.
\newblock Psychology \& Marketing \textbf{24}(11), 975--1000 (2007)

\bibitem{Dabbish:2012}
Dabbish, L., Stuart, C., Tsay, J., Herbsleb, J.: Social coding in github:
  Transparency and collaboration in an open software repository.
\newblock In: CSW12, pp. 1277--1286. ACM (2012)

\bibitem{Datta2015}
Datta, S., Bhatt, D., Jain, M., Sarkar, P., Sarkar, S.: The importance of being
  isolated: An empirical study on chromium reviews.
\newblock In: ACM/IEEE ESEM, pp. 1--4 (2015)

\bibitem{Ford2017b}
Ford, D.: Using eye tracking to identify features of peer parity on stack
  overflow.
\newblock In: VL/HCC, pp. 319--320 (2017)

\bibitem{Ford2017a}
Ford, D., Harkins, A., Parnin, C.: Someone like me: How does peer parity
  influence participation of women on stack overflow?
\newblock In: VL/HCC, pp. 239--243 (2017)

\bibitem{Gershenson2017}
Gershenson, S., Hart, C.M.D., Lindsay, C.A., Papageorge, N.W.: The Long-Run
  Impacts of Same-Race Teachers (2017)

\bibitem{G13}
Gousios, G.: The {GHT}orrent dataset and tool suite.
\newblock In: MSR, pp. 233--236 (2013)

\bibitem{Anders2013}
Larsson, A.O., Ihlen, A.: Birds of a feather flock together? party leaders on
  twitter during the 2013 norwegian elections.
\newblock European Journal of Communication \textbf{30}(6), 666--681 (2015)

\bibitem{Lee2017}
Lee, R.K.W., Lo, D.: Github and stack overflow: Analyzing developer interests
  across multiple social collaborative platforms.
\newblock In: SocInfo (2017)

\bibitem{Lima2014}
de~Lima, A.M.G., Rossi, L., Musolesi, M.: Coding together at scale: Github as a
  collaborative social network.
\newblock CoRR  (2014)

\bibitem{Lu11576}
Lu, X., Bengtsson, L., Holme, P.: Predictability of population displacement
  after the 2010 haiti earthquake.
\newblock pp. 11,576--11,581. National Academy of Sciences (2012)

\bibitem{Majumder2012}
Majumder, A., Datta, S., Naidu, K.: Capacitated team formation problem on
  social networks.
\newblock In: KDD, pp. 1005--1013 (2012)

\bibitem{Mamykina:2011}
Mamykina, L., Manoim, B., Mittal, M., Hripcsak, G., Hartmann, B.: Design
  lessons from the fastest q\&a site in the west.
\newblock In: CHI, pp. 2857--2866. ACM (2011)

\bibitem{Pappalardo2015}
Pappalardo, L., Pedreschi, D., Smoreda, Z., Giannotti, F.: Using big data to
  study the link between human mobility and socio-economic development.
\newblock In: International Conference on Big Data (Big Data), pp. 871--878.
  IEEE (2015)

\bibitem{immigPaper}
Rostila, M.: Birds of a feather flock together – and fall ill? migrant
  homophily and health in sweden.
\newblock Sociology of health and illness (32) (2010)

\bibitem{Sarma:2007}
Sarma, A., Bortis, G., van~der Hoek, A.: Towards supporting awareness of
  indirect conflicts across software configuration management workspaces.
\newblock In: ASE, pp. 94--103. ACM (2007)

\bibitem{Sharma2016}
Sharma, A., Chen, X., Kuttal, S., Dabbish, L., Wang, Z.: Hiring in the global
  stage: Profiles of online contributions.
\newblock In: ICGSE (2016).
\newblock \doi{10.1109/ICGSE.2016.35}

\bibitem{Subramaniam2012}
Subramaniam, M.M., Ahn, J., Fleischmann, K.R., Druin, A.: Reimagining the role
  of school libraries in stem education:creating hybrid spaces for exploration.
\newblock The Library Quarterly \textbf{82}(2), 161--182 (2012)

\bibitem{ImmigrantBlog}
Szebenyi, D.:
  \url{https://www.liberties.eu/en/infographics/migration-in-the-eu-infographic/112}.
\newblock Online Accessed: 2016-01-22

\bibitem{Tang2013}
Tang, L.: Does ``birds of a feather flock together'' matter - evidence from a
  longitudinal study on us-china scientific collaboration.
\newblock Journal of Informetrics \textbf{7}(2), 330 -- 344 (2013)

\bibitem{Thung2013}
Thung, F., Bissyande, T.F., Lo, D., Jiang, L.: Network structure of social
  coding in github.
\newblock In: CSMR, pp. 323--326. IEEE (2013)

\bibitem{Tsay:2014}
Tsay, J., Dabbish, L., Herbsleb, J.: Influence of social and technical factors
  for evaluating contribution in github.
\newblock In: ICSE, pp. 356--366 (2014)

\bibitem{Hamm2000}
V.~Hamm, J.: Do birds of a feather flock together? the variable bases for
  african american, asian american, and european american adolescents'
  selection of similar friends \textbf{36}, 209--19 (2000)

\bibitem{Vasilescu2013}
Vasilescu, B., Filkov, V., Serebrenik, A.: Stackoverflow and github:
  Associations between software development and crowd sourced knowledge.
\newblock In: 2013 SocialCom, pp. 188--195 (2013)

\bibitem{SNA}
Wasserman, S., Faust, K.: Social network analysis: Methods and applications
  (1994)

\bibitem{Yu:2014}
Yu, Y., Yin, G., Wang, H., Wang, T.: Exploring the patterns of social behavior
  in github.
\newblock In: CrowdSoft, pp. 31--36 (2014)

\end{thebibliography}
